 \documentclass[final,5p,times,twocolumn,authoryear]{elsarticle}

\usepackage{amssymb}
\usepackage{amsmath}

\usepackage{graphicx}
\usepackage{dcolumn}
\usepackage{CJK}
\usepackage[german, english]{babel}
\usepackage{bm}
\usepackage{aasmacros}
\usepackage{doi}
\usepackage{xcolor}
\usepackage{tikz}
\definecolor{orcidlogocol}{HTML}{A6CE39}

\defcitealias{Kroupa+2023}{Kroupa, Gjergo, \emph{et al.}, (2023)}
\defcitealias{Kroupa+2024}{Kroupa, Gjergo, Jerabkova \& Yan (2024)}
\newcommand{\diff}[0]{\mathrm{d}}

\definecolor{burntorange}{rgb}{1., 0.349, 0.0}

\journal{Nuclear Physics B}

\begin{document}

\begin{frontmatter}



\title{The Impact of Early Massive Galaxy Formation on the Cosmic Microwave Background}


\author[first,second]{Eda Gjergo}
\ead{eda.gjergo@gmail.com}
\affiliation[first]{organization={School of Astronomy and Space Science, Nanjing University},
            city={Nanjing},
            postcode={210093}, 
            state={Jiangsu},
            country={People’s Republic of China}}
\affiliation[second]{organization={Key Laboratory of Modern Astronomy and Astrophysics, Nanjing University, Ministry of Education},
            city={Nanjing},
            postcode={210093}, 
            state={Jiangsu},
            country={People’s Republic of China}}

 \author[label1,label2]{Pavel Kroupa}
 \ead{pkroupa@uni-bonn.de}
\affiliation[label1]{
organization={Helmholtz-Institut f\"ur Strahlen- und Kernphysik, Universit\"at Bonn},
            addressline={Nussallee 14-16}, 
            city={Bonn},
            postcode={53115}, 
            state={North Rhine-Westphalia},
            country={Germany}}
\affiliation[label2]{organization={Astronomical Institute
  Charles University},
            addressline={V Holesovickach 2},
            city={Praha},
            postcode={18000},
            state={Bohemia},
            country={Czech Republic}}

\begin{abstract}
The Cosmic Microwave Background (CMB) anisotropies, corrected for foreground effects, form the foundation of cosmology and support the Big Bang model. 
A previously overlooked foreground component is the formation of massive early-type galaxies (ETGs), which can no longer be ignored, particularly in light of JWST’s detection of massive, evolved systems at extreme redshifts ($z > 13$). The rapid formation of massive ETGs has been advocated in galaxy evolution studies for decades, and recent evidence has compelled even proponents of hierarchical mass assembly to acknowledge the fact that massive ETGs evolve quickly. Constraints from chemical evolution are particularly stringent. Without both intense star~formation and a top-heavy galaxy-wide initial~mass~function of stars (IMF), it is difficult to reconcile stellar~population~synthesis models with the high metallicity and abundance patterns of $\alpha$-elements. 
We infer from previous studies that the progenitor cloud of each massive ETG must have had a radius of $\approx 400$~kpc. Comparing this value to the average present-day separation of massive ETGs, their formation may have occurred around $15<z<20$. 
We consider this epoch of formation in a flat-$\Lambda$CDM cosmological context, incorporating the known and necessary properties of massive ETGs -- properties encapsulated independently by the integrated~galaxy-wide~IMF (IGIMF) theory. 
The massive ETG evolution presented in this work is consistent with recent advancements in stellar and galaxy evolution, and is derived entirely without priors or constraints from the CMB. Yet, it emerges as a non-negligible source of CMB foreground contamination. 
Even in our most conservative estimates, massive ETGs account for $1.4$\% up to the full present-day CMB energy density. 
\end{abstract}



\end{frontmatter}



\section{Introduction\label{sec:intro}}

The Cosmic Background Explorer  \citep[COBE, ][]{Mather+1990} was the first satellite mission to confirm that the Cosmic Microwave Background (CMB) is a near-perfect blackbody with a temperature of 2.725 K. To achieve this result, it was necessary to accurately account for both instrumental systematic errors and foreground contamination, primarily from Galactic synchrotron and dust emission \citep{Smoot+1990}. The root-mean-square fluctuations in the CMB temperature are of the order of $\Delta T / T \approx 10^{-5}$ \citep{Fixsen+1996}. 
This near-isotropy has been a significant triumph for the Big Bang model, which predicts that the afterglow of recombination should produce a blackbody spectrum -- provided that primordial quantum fluctuations were amplified by inflation through 60~$e$-folds nearly instantaneously. 

After COBE, other explanations to the CMB fell out of favor. For example, if multiple sources contributed to the observed CMB, each emitting blackbody radiation with their own temperature and redshift distributions, their combined radiation would not generally sum to a single blackbody
. Only an unlikely alignment of source properties, such as the redshift-scaling of the blackbody temperature by ($z+1$) would lead to a cumulative radiation field that still appears both in shape and in intensity as a blackbody \citep[e.g.,][p. 289]{Peacock1999}. COBE hence paved the way to precision cosmology, a discipline reliant upon the investigation of the CMB anisotropies to constrain cosmological parameters.

Recent efforts in CMB science have been geared toward foreground subtractions -- mainly of Galactic dust foreground -- in order to understand the CMB polarization \citep[e.g.,][]{Pogosian+2019Future}, but also on proper extragalactic foregrounds below $z<2$ \citep{MacCrann+2024AtacamaCMBforegrounds}. In the foreground subtraction of extragalactic sources, however, a possible component has been neglected, namely the early formation of massive galaxies. In particular, we consider here massive (with a stellar mass of $M_{\star}> 10^{11.5} M_{\odot}$) early-type galaxies (ETGs). For the purposes of this work, we use ``ETGs" interchangeably with elliptical galaxies, which are ellipsoidal smooth and dense structures whose present-day stellar population is primarily old and metal-rich. 

Historically, two competing models have been proposed to explain galaxy formation (including ETGs): the hierarchical scenario and the monolithic scenario. In the hierarchical scenario, small structures form first and merge over time to build progressively larger systems, such as galaxies and galaxy clusters. This scenario arises naturally in cold dark matter cosmologies, where the initial matter power spectrum is dominated by small-scale fluctuations \citep[e.g.,][]{Peebles1974}. The monolithic scenario, by contrast, holds that compact structures, such as elliptical galaxies and bulges, form early through the rapid collapse of a large gas cloud, followed by a starburst phase \citep[e.g.,][]{Eggen+1962}. Recent observations with the James Webb Space Telescope (JWST) have uncovered massive quiescent galaxies with stellar masses of more than $10^{11}\,M_\odot$ that formed within 1--2~Gyr after the Big Bang \citep[e.g., ][]{Glazebrook+2024}. JWST data have also uncovered massive galaxies forming at redshifts, $z\approx 15$ \citep[e.g.,][and references therein]{Haslbauer+2022b, McGaugh+2024}, strongly supporting the monolithic scenario rather than a hierarchical buildup through mergers, as was shown by \citet{Haslbauer+2022b} and most recently by \citet{McGaugh+2024}. The inferred redshifts of JWST-detected galaxies continue to increase as data analysis techniques and photometric constraints improve \citep{Castellano+2025}. In conjunction with data from the Alma Large Millimeter Array (ALMA), these observations possibly indicate strong dust obscuration of the star-forming galaxies \citep[e.g.]{Schreiber+2018, Bakx+2024, Langeroodi+2024, Nikopoulos+2024}. 

Originally proposed by \citet{Eggen+1962}, the monolithic scenario has found support over subsequent decades \citep{MatteucciBrocato1990, ChiosiCarraro2002, Renzini2006}. Adjacent is the ``\emph{impossibly early galaxy problem}''\citep{Steinhardt+2016}, wherein cold~dark-matter-based hierarchical mass assembly predicts that the transition from  halo formation to baryonic evolution should occur predominantly at a redshift range of $4<z<8$. Yet, massive halos of $10^{12-13}M_{\odot}$ have already been observed at $6<z<8$. Stellar population studies further indicate that the more massive an elliptical galaxy, and the denser its environment, the earlier and shorter its formation timescale \citep{Thomas+2005}. This phenomenon is called ``\emph{downsizing}". Astrophysical probes appear to contradict the purely dark-matter-driven hierarchical formation model. 
Even within the hierarchical structure formation framework, intricate dynamical and cooling mechanisms have been introduced to support a two-phase galaxy formation scenario, where the bulk of massive ETGs forms rapidly \citep[e.g.,][]{Mo+2024}.

Arguments for the downsizing scenario find strong support in chemical evolution. Decades ago, it was demonstrated \citep{Sandage1986, MatteucciBrocato1990, PipinoMatteucci2004} that the observed abundance patterns of elements are incompatible with a purely hierarchical mass assembly framework. Specifically, there is a well-established positive correlation between [$\alpha$/Fe] abundances\footnote{The square bracket notation represents the normalization of the ratio of two species, $A$ and $B$, either by number or mass, relative to solar values $\odot$: $[A/B] = \log \left(\frac{N_A}{N_B}\right) - \log \left(\frac{N_A}{N_B}\right)_{\odot}$.} and the masses of early-type galaxies (ETGs) \citep{Thomas+2002}. High $\alpha$~abundances at high metallicities indicate rapid enrichment by massive stars. 
\citet{Yan+2021} has recently shown that the most natural way to reconcile most observational probes -- from stellar population synthesis models, to downsizing, as well as Type Ia supernova rates and $\alpha$-element abundance patterns -- is to implement a top-heavy, galaxy-wide initial mass function of stars (IMF)\footnote{The {\it galaxy-wide IMF} describes the mass distribution of all the stars a galaxy forms during a short time interval (e.g., 10 Myr). It differs from the {\it stellar IMF}, which instead refers to the IMF of a single star-forming event.}. This means that the distribution of stars formed in massive ETGs must have been skewed toward the preferential production of massive stars.

 Another, entirely independent pathway to quantifying the formation rate of massive stars in these forming ETGs is through analysis of star counts in stellar-dynamically bound populations such as globular star clusters and ultra-compact dwarf galaxies (UCDs). The early phases of UCDs can be recovered by combining information from present-day starburst systems with knowledge on stellar-dynamical processes \citep{Marks+2012}. This analysis  suggests that systems with a high star-formation rate densities produce distributions with an overabundance of massive stars compared to the local star formation activity observed in the vicinity of the Sun within the Galaxy.

The galaxy-wide mass function of newly-formed stars is determined by integrating the distributions of stellar populations formed over time. A robust empirical framework describing the galaxy-wide IMF is provided by the Integrated Galaxy-wide Initial Mass Function (IGIMF) theory \citep{KroupaWeidner2003, Jerabkova+2018, Kroupa+2024}. This theory is based on the simple concept that stars form in molecular cloud overdensities, i.e. clumps which spawn embedded clusters. Each embedded cluster forms a population of stars, the mass-distribution of which follows the stellar initial mass function (IMF) as formulated by \cite{Marks+2012}. The addition of all these forming embedded clusters -- i.e., the integration over the whole forming galaxy -- leads to the galaxy-wide IMF of stars, or the IGIMF. The IGIMF depends, at any time, primarily on the star-formation rate (SFR) of the galaxy, which is fundamentally determined by the mass-spectrum of dense gas cloud clumps in it, but also on the metallicity of the gas, which dictates its ability to cool and consequently, to fragment. Given a star-formation history (SFH) of a galaxy, we can therefore calculate the complete spectrum of all stars ever formed in the galaxy from first principles. Applications of the IGIMF theory have successfully resolved several previously unexplained extragalactic phenomena. These include the stellar-mass–metallicity relation of galaxies, the discrepancy between extended galactic UV disks and their smaller H$\alpha$ radii, the downsizing phenomenon, and the high dynamical mass-to-light ratios observed in ultra-compact dwarf galaxies and massive ETGs. For a comprehensive review, see \citetalias{Kroupa+2024}.

These results -- derived from resolved stellar populations showing variations in the stellar IMF with the physical conditions of star-forming gas -- are in excellent agreement with observational constraints on the variation of the galaxy-wide IMF shape with galaxy SFRs \citep{Yan+2017}. Specifically, from low-mass late-type galaxies with SFR$\,\approx 10^{-5}\,M_\odot/$yr, the galaxy-wide IMF is observed to lack massive stars \citep[e.g.,][]{Watts+2018}. As the SFR increases, the relative number of massive stars also increases  \citep{Lee+2009},  eventually leading to a top-heavy galaxy-wide IMF with a significant surplus of massive stars in massive late-type galaxies exhibiting SFRs of $\,\approx 50\,M_\odot/$yr \citep{Gunawardhana+2011}. The observational data and the calculated IGIMF results are explicitly compared in Fig.~6 of \cite{Yan+2017}. 

Furthermore, the variation of the galaxy-wide IMF as calculated using the IGIMF theory leads naturally, in combination with the fast formation of ETGs and galaxy bulges,  to the rapid birth of super-massive black holes and to their correlation with the host-galaxy properties \cite{Kroupa+2020}. The IGIMF also naturally converges, as stated earlier, with the variation of the galaxy-wide IMF required to explain the very rapid nucleosynthesis responsible for the copious chemical enrichment observed in massive ETGs \cite{Yan+2021}. 

Of particular relevance to understanding the physical mechanism behind the creation of the CMB are the observed deviations from perfect isotropy and homogeneity across the sky. 
While the inflation-driven Hot Big Bang theory, a cornerstone of the $\Lambda$CDM model, predicts homogeneity and isotropy, numerous studies \citep{Eriksen+2004, Schwarz+2004, Eriksen+2007, Jones+2023} have demonstrated significant tensions when the CMB power spectrum is interpreted solely within this framework.  
One of these tensions is the lack of correlation found for angular scales larger than about 60~degrees, suggesting that the very early Universe was not causally connected on these scales, in contradiction to the inflation-driven Hot Big Bang model \citep[e.g.,][]{Schwarz+2016}. Perhaps more directly relevant to the problem at hand, the southern celestial hemisphere exhibits more power in the CMB than the northern hemisphere, an anisotropy that is incompatible with the inflation-driven Hot Big Bang model. \citet{Kroupa+2023} noted that the southern hemisphere also contains a significantly larger number of early-type galaxies \citep{JavanmardiKroupa2017}. This suggests a potential connection between the CMB and ETGs.

Our  rationale is as follows: we are immersed in the CMB photon energy density field, $U_{\rm CMB}(z=0)$, whose present-day value scales with the fourth power of the CMB temperature (Eq.~\ref{eq:SBEdens-z}).  At earlier times, this energy density was higher, scaling as $(1+z)^4$. The CMB photon field has traversed the entire observable Universe to reach us, interacting with several foreground sources -- including the birth of massive ETGs. As outlined in the introduction, stringent constraints suggest that massive ETGs formed very early, very rapidly, and with a top-heavy galaxy-wide IMF -- that is, with a high-mass slope shallower than that observed in present-day star-forming galaxies. Consequently, their luminosity at formation must have been significantly higher than they are today, when they are powered by long-lived, low-mass stars. The top-heavy galaxy-wide IMF also implies a rapid onset of chemical enrichment. Given that chemical enrichment is the bottleneck to dust formation (see Sec.~\ref{sec:method-Edens-ETGdust}), we implement the approximation that the intense luminosity from star-forming ETGs was in thermal equilibrium with dust. If this emission was processed efficiently, the resulting photon energy density field from massive ETGs could have rivaled that of the CMB.

With this contribution, we assume standard cosmology and we adopt the currently accepted Planck cosmological solution \citep{Planck2018}. We then include the observational information pertaining to the rapid, very early formation of massive ETGs to investigate their possible role in an apparent CMB contribution.  
The method (Sec.~\ref{sec:method}) is organized as follows: we first construct the luminosity evolution of an average massive ETG, based on existing knowledge from galaxy and stellar evolution (Sec.~\ref{sec:method-singleETG}). Next, we develop a birthrate function for all massive ETGs as a function of cosmic time (Sec.~\ref{sec:method-ETGcosmoVol}). We then derive the evolution of the 
energy density in photons generated by such sources
(Sec.~\ref{sec:method-Edens}). In the results (Sec.~\ref{sec:results}) we examine the properties that emerge from the considerations from Sec.~\ref{sec:method}, and we discuss their implications in Sec.~\ref{sec:discussion}, placing them in the context of the historical development of the established CMB cosmology. We provide our conclusions in Sec.~\ref{sec:conclusions}, where we explore the potential consequences arising from this work.

\section{Method\label{sec:method}}

In this section we present the average bolometric luminosity evolution of massive ETGs, we construct their birth~function, and then convolve these quantities within a cosmological volume. Next,  we explain why this luminosity is expected to be fully thermalized by dust. We derive the massive ETG present-day contribution to the background radiation, adopting a flat $\Lambda$CDM cosmological framework (unless stated otherwise), using parameters from \citet{Planck2018}. 
Lastly, we compare the evolution of the total 
energy density
produced by dust-processed massive ETG radiation to the evolution of the present-day CMB black~body.

\subsection{Characterizing massive ETG sources}\label{sec:method-singleETG}
\subsubsection{Characterizing the properties of massive ETGs}\label{sec:method-singleETG-ETGproperties}

Our working hypothesis, consistent with  downsizing \citep{Thomas+2005}, as well as with recent observations \citep{McGaugh+2024}, is that the earliest galaxies to form are also the most massive ETGs (with present-day stellar masses $M_*^{\rm today}> 10^{11.5} M_{\odot}$). 
From  stellar population observations \citep[e.g.,][]{Idiart+2007Ages, Lacerna+2020SDSS-IV}, N-body simulations \citep[e.g.,][]{ChiosiCarraro2002}, and chemical evolution studies \citep[e.g.,][]{PipinoMatteucci2004}, the evolution of elliptical galaxies must occur with energetic near-monolithic starbursts spanning timescales of a few hundred mega-years (Myr). \citet{Merlin+2012} is one of such studies where the SFR associated with the formation of massive ETGs must be of the order of $10^3 M_{\odot}/\text{yr}$. From chemical evolution models in particular, there are stringent constraints on the galaxy-wide IMF that describes the mass distribution of stars within self-contained star~formation episodes. For low iron abundances, associated with earlier epochs when the metallicity was lower, the distribution of $\alpha$-elements \footnote{Multiples of He, $\alpha$-elements are those whose atomic number is even, e.g., O, Mg, Si, through Ca.} correlates with the galaxy mass and morphology \citep[e.g.,][]{MatteucciTornambe1987, MatteucciBrocato1990}, with more massive ETGs containing the most $\alpha$~enhancement  -- tell-tale evidence for their very rapid formation. 

More recently, \citet{Yan+2021} unified, for the first time, several independent theoretical elements and reconciled them with observations of ETGs. Their theoretical elements include formation timescales within the downsizing scenario, stellar population synthesis models, Type Ia supernova rates, and chemical evolution. Among the key observables are the dominance of M-dwarf stars in the stellar populations of massive ETGs, their large dynamical mass-to-light ratios, and the characteristic $\alpha$-to-Fe abundance patterns. It is the short formation timescales of massive ETGs, the rapid accumulation of nucleosynthesis products, and the dependence of the low-mass stellar IMF on metal abundance that drive the present-day prevalence of M-dwarf stars in these galaxies. These properties are well encapsulated by the IGIMF theory \citep{Kroupa+2024}, and, most intriguingly, align with the same phenomenological formulation of the IGIMF derived from ultra-compact dwarf galaxies (UCDs), as demonstrated by \citet{Marks+2012}. Independently, also \citet{BekkiTsujimoto2023} found that UCDs require a top-heavy stellar IMF.
In summary, massive ETGs formed rapidly through intense starbursts ($\text{SFR}\approx10^3 M_{\odot}/{\text{yr}}$). We take $10^{11.5}$ and $10^{12} M_{\odot}$ to be the upper and lower limit for the present-day ETG stellar masses ($M_{*, \text{ETG}}^{\text{today}}$). 

\subsubsection{Timescales of formation of massive ETGs}\label{sec:method-singleETG-ETGtimescale}

Instead of $M_{*, \text{ETG}}^{\text{today}}$, let us now consider $M_{*,\text{ETG}}^{\text{tot}}$, the cumulative stellar mass produced by ETGs over their lifetimes (from their epoch of formation, $t_{\rm form} \text{, to the present day, } t_{\rm now} $):
\begin{equation}\label{eq:totcumstellarmass}
M_{*,\text{ETG}}^{\text{tot}} = \int_{t_{\rm form}}^{t_{\rm now}} \text{SFR}(t) \, \mathrm{d}t \, .
\end{equation}
Using the downsizing formulation from \citet{Recchi+2009} (their Fig.~18), formalized by \citet{Kroupa+2020} (their Eq.~1)\footnote{Note that in \citet{Kroupa+2020}, $M_{*,\text{ETG}}^{\text{tot}}$ was called $M_{\rm igal}$, i.e., the total stellar mass of the galaxy which initially participated in the monolithic collapse.}:
\begin{equation}
    \tau_{\rm down} = 8.16 {\rm e}^{(-0.556 \log_{10}(M_{*,\text{ETG}}^{\text{tot}}/M_{\odot})+3.401)} + 0.027 \, .
\end{equation}
We find that the formation timescales of ETGs, $\tau_{\text{down}}$,  are given by:
\begin{equation}
\tau_{\text{down}}(M_{*,\text{ETG}}) =
\begin{cases}
440 \, \text{Myr}, & \text{if } M_{*,\text{ETG}}^{\text{tot}} =  10^{11.5} \, M_\odot \, , \\
340 \, \text{Myr}, & \text{if } M_{*,\text{ETG}}^{\text{tot}} = 10^{12} \, M_\odot \,.
\end{cases}
\end{equation}
In the examples given, we picked again the upper and lower limit of $M_{*, \text{ETG}}^{\text{today}}$, i.e., $10^{11.5} \, M_{\odot} \text{ and } 10^{12} \, M_{\odot}$. Note however that $M_{*, \text{ETG}}^{\text{today}}$ is necessarily smaller than $M_{*,\text{ETG}}^{\text{tot}}$, because it excludes the mass from stars with lifetimes shorter than the present-day age of the galaxy.
Concerning the relationship between $M_{*,\text{ETG}}^{\text{tot}}$ and $M_{*, \text{ETG}}^{\text{today}}$, we note that under a canonical and invariant form of the galaxy-wide IMF \citep{Kroupa2001} and assuming a flat SFR lasting about 1 Gyr, massive ETGs will lose about 30~per~cent of their cumulative stellar mass by the present epoch (i.e., $M_{*, \text{ETG}}^{\text{today}} \approx 0.7 \,M^{\rm tot}_{*, \text{ETG}}$, \citealt{BaumgardtMakino2003}, e.g. their Fig.~1), because all stars forming in the first main starburst of the massive ETG will have died by the present time \citep[for estimates on stellar lifetimes, see, e.g.,][]{portinari+1998}. Under the IGIMF theory, this estimate of mass loss of roughly 0.3 goes up to about $0.99$ \citep[see Fig.~3 and Fig.~4 of ][ respectively]{Kroupa+2020, Mahani+2021}. This means that the $M_{*,\rm ETG}^{\rm tot}$ of the most massive ETGs could have been as high as $10^{14} M_{\odot}$. As we have noted earlier (last paragraph of Sec.~\ref{sec:method-singleETG-ETGproperties}) it is only with the IGIMF theory that all observational and theoretical aspects of ETGs can find coherence.

\subsubsection{Average separation of massive ETGs}\label{sec:method-singleETG-ETGdistance}
In the local Universe, the average separation of  massive ETGs is $\langle d_0 \rangle \approx 15$~Mpc \citep{Davies+1993}. 

In order to reproduce the formation timescales of ETGs deduced by \cite{Thomas+2005}, i.e. the downsizing timescales, gravitationally collapsing progenitor gas clouds had to have had a radius of about $\langle r_{f}\rangle \approx 400\,$kpc as shown by \citet[][their Fig.~4]{Eappen+2022}.

At the same time, from \citet[][their Fig.~4]{Eappen+2022} we  find that such ETGs should be forming from the gravitational collapse of a progenitor gas cloud whose radius at the time of formation was $\langle r_{f}\rangle \approx 400$~kpc. 

For any cosmological model based on the Friedmann-Lema\^{i}tre-Robertson-Walker (FLRW) geometry, the following relation must hold:
\begin{align}
    1+\langle z_{\rm f}\rangle &= \frac{\langle d_0 \rangle}{\langle d_{f} \rangle}\\
    \langle z_f \rangle &= \frac{15 \text{Mpc}}{800 \text{kpc}} - 1 \approx 16.5 \, ,
\end{align}
which links the average redshift of formation $\langle z_f \rangle$ to the average present-day separation of massive ETGs  and average separation at the epoch of formation, $\langle d_f\rangle = 2 \langle r_f \rangle \approx 800 \text{ kpc}$.
In a valid cosmological model, this redshift, $\langle z_f\rangle \approx 16.5$, should identify an epoch encompassing both an average galaxy formation timescale $\langle \tau_{\text{down}}\rangle$ as well as the formation timescale of the parent clouds.

\subsubsection{Possible connection with the 21-cm anomaly}\label{sec:method-singleETG-21cm}
Very close to our estimate $\langle z_f \rangle \approx 16.5$ is the 21-cm anomaly \citep{Bowman+2018-21-cm-anomaly}, centered at a redshift of $z=17.5$ and whose dip in brightness temperature spans a redshift range of about $15 < z< 20$. While exotic physics has been invoked to explain this anomaly, its astrophysical explanation could be simpler: either there exists an excess radio background which enhances the contrast between the CMB and the gas temperature, or there is some other mechanism which cools gas efficiently at this epoch. We consider this second scenario, which is not widely accepted given that standard star~formation conditions would cause a depression in the signal only half as deep \citep{MirochaFurlanetto2019}. Best fits to the 21-cm anomaly can only be achieved with extreme star~formation~efficiency paired with the presence of top-heavy galaxy-wide IMFs of metal-free stars \citep{MittalKulkarni2022}. Incidentally, these conditions coincide with the independent models explained in Sec.~\ref{sec:intro} and Sec.~\ref{sec:method-singleETG-ETGproperties} concerning the massive ETG properties.

\subsubsection{Birth function of SFUs}\label{sec:singleETG-birthfunction}

The full, self-consistent luminosity evolution of an ETG can be modeled using advanced codes that reconstruct the stellar population and chemical evolution within a galaxy, such as {\tt photGalIMF}\footnote{https://github.com/juzikong/photGalIMF} \citep{Haslbauer+2024}. In this work, we introduce a simplified method that provides a first-order approximation in good agreement with fully self-consistent calculations. 

In particular, we adopt the luminosity evolution $\mathcal{L}_{\text{SFU}}(t)$ from \citet[][ e.g. their Fig.~7]{Jerabkova+2017}, for a massive star-forming unit (SFU) evolving according to the IGIMF, whose present-day bolometric luminosity is several $10^6 L_{\odot}$, with a present-day V-band mass-to-light ratio $M/L\approx10$ 
\citep[see Fig.~1 and Fig.~6 of ][ respectively]{Kroupa+2020, Mahani+2021}, in Solar units, where $M$ encompasses the mass in all stars and their remnants. Such an object would have had a cumulative star~formation yield of $M_{*,\text{SFU}}^{\rm tot}=\int_{t_{\text{form}}}^{t_{\text{now}}} \text{SFR}(t) \diff t = 10^9 M_{\odot}$ within an IGIMF framework (see Sec.~\ref{sec:method-singleETG-ETGtimescale}). This evolution is reported on the top~panel of Fig.~\ref{fig:singleETG}. We take the mass distribution of SFUs in our ETG to be constant. The stellar~birth~function of our ETG is given by a normal distribution, symmetric with respect to linear time -- the ETG age -- where the downsizing timescale is equivalent to the full-width-half-max (FWHM) of the normal distribution. The standard deviation $\sigma_{\text{SFU}}$ of this distribution is therefore described by:
\begin{equation}
    \sigma_{\text{SFU}} = \frac{\tau_{\text{down}}}{2\sqrt{2\ln(2)}} \, .
\end{equation}
We center the normal distribution $\mathcal{B}_{\text{SFU}}$ describing the unitary stellar~birth~function around an average time of formation $\mu_{\text{SFU}}= 2\, \tau_{\text{down}}$,
The factor 2 is effectively a buffer between the origin (zeroth time) and the onset of galaxy formation.

\begin{equation}
    \mathcal{B}_{\text{SFU}}(t_{\text{ETG}}) = \frac{1}{\sqrt{2\pi\sigma^2_{\text{SFU}}}}e^{-\frac{(t_{\text{ETG}}-\mu_{\text{SFU}})^2}{2\sigma_{\text{SFU}}^2}} \, . 
    \label{eq:SFU-birthrate}
\end{equation}
The distribution $N_{\text{SFU}}\,\mathcal{B}_{SFU}(t_{\text{ETG}}) = \diff N_{\text{SFU}}/\diff t_{\text{ETG}}$, where $N_{\text{SFU}}$ represents the number of SFUs that will form during the birth of an average massive ETG, is shown in the middle panel of Fig.~\ref{fig:singleETG}. In reality, the stellar~birth~distribution is unlikely to be a Gaussian, symmetric w.r.t. the linear ETG age. But instead of arbitrarily parametrizing gamma or log-normal distributions, we defer a more realistic calculation of the early epochs to detailed studied with {\tt photGalIMF} \citep{Haslbauer+2024} paired with the collapse simulations of \citet{Eappen+2022}. 

\subsubsection{Luminosity evolution of an average ETG}\label{sec:method-singleETG-LumEvol}

From the calculations implemented in the previous section, the resulting peak in the stellar~birth~function occurs at  $\approx$ 259 Myr (also seen in Fig.~\ref{fig:singleETG}), corresponding to a redshift of $z\approx 15.5$ under the unrealistic assumption that the ETG began forming right after the Big Bang. However, this conversion was performed within the framework of flat-$\Lambda$CDM. Observations of mature, evolved galaxies at high redshifts by JWST provide the latest to a conspicuous body of evidence \citep[e.g.,][]{Kroupa+2023, DiValentino2022, Steinhardt+2016},
suggesting that the current standard model of cosmology will not remain viable in its present form for much longer.
It has already been documented that the time-line for forming the observed massive elliptical galaxies is not possible in a $\Lambda$CDM universe \cite{Thomas+2005, Eappen+2022}. While they must have formed rapidly in reality, in  $\Lambda$CDM they need billions of years to assemble. 
Nonetheless, while the time-to-redshift conversion exhibits some inconsistencies, it serves as a reasonable approximation for the purposes of this study. The results presented here highlight an existing problem that future cosmological models must address.

For an average massive ETG, its bolometric luminosity evolution $\mathcal{L}_{\text{ETG}}(t_{\text{ETG}})$ as a function of ETG age, $t_{\text{ETG}}$, will be given by the convolution of its stellar birth function, $\mathcal{B}_{\text{SFU}}(t)$, with the bolometric luminosity of its star-forming units, $\mathcal{L}_{\text{SFU}}$:
\begin{equation}\label{eq:LsingleETG}
    \mathcal{L}_{\text{ETG}}(t_{\text{ETG}}) = \int_{t_f}^{t_0}N_{\text{SFU}} \, \mathcal{B}_{\text{SFU}}(t') \, \mathcal{L}_{\text{SFU}}(t_{\text{ETG}} - t')\,  \diff t \, .
\end{equation}
The results for an average massive ETG whose present-day mass is $M_{*,\text{ETG}}^{\rm today}=10^{12}M_{\odot}$ are reported on the bottom panel of Fig.~\ref{fig:singleETG}.

\subsection{Evolution of ETGs in a cosmological volume}\label{sec:method-ETGcosmoVol}
\subsubsection{Cosmic ETG birth function}\label{sec:method-ETGcosmoVol-ETGbirthfunction}
Numerous observational issues \citep[e.g.,][]{Kroupa+2023, McGaugh+2024} hint at the fact that gravity under weak-field regimes, including structure~formation conditions, is poorly understood. We therefore proceed with the ansatz (whose motivations are explained in Sec.~\ref{sec:method-singleETG-21cm}) that the 21-cm anomaly coincides with the peak in star~formation for massive ETGs. 
To convert to cosmic age, $t_{\text{cosmo}}$, from redshift, we use \citep[e.g.,][their Eq.~30]{Hogg1999}:
\begin{equation}
    t_{\text{cosmo}}(z)= \int_z^{\infty}\frac{1}{(1+z')H(z')}\diff z' \, ,
\end{equation}
where we take the Hubble parameter to be $H(z) = H_0 \sqrt{\Omega_{m}(1+z)^3+\Omega_{\Lambda}}$ and $H_0=67.66\, \text{km/s}\,\text{Mpc}^{-1}$ is the Hubble constant from \citet{Planck2018}. By $z=20$ we are out of the radiation-dominated regime, so in flat-$\Lambda$CDM we can ignore any energy density component other than matter and dark energy.

The redshifts of 20 and 15 will therefore correspond to times $t_{\mathcal{B}(\text{ETG,i})}$ and $t_{\mathcal{B}(\text{ETG,f})}$ (refer to the limits of Eq.~\ref{eq:LsingleETG}), respectively, which we take to contain $\pm 3\sigma$ of all the ETG births. 
The resulting ETG~birth~function for a comoving volume will hence be given by a normal distribution, symmetric w.r.t. linear cosmic time:
\begin{align}
    \mu_{\text{ETG}} &= \frac{t_{\mathcal{B}(\text{ETG,i})}+t_{\mathcal{B}(\text{ETG,f})}}{2}\\
    \sigma_{\text{ETG}} &= \frac{|t_{\mathcal{B}(\text{ETG,f})}-t_{\mathcal{B}(\text{ETG,i})}|}{6}\\
    \mathcal{B}_{\text{ETG}}(t_{\text{cosmo}}) &= \frac{1}{\sqrt{2\pi\sigma^2_{\text{ETG}}}}e^{-\frac{(t_{\text{cosmo}}-\mu_{\text{ETG}})^2}{2\sigma_{\text{ETG}}^2}} \, . 
    \label{eq:ETG-birthrate}
\end{align}

To integrate these birth~rates ($\mathcal{B}_{\text{SFU}}$, Eq.~\ref{eq:SFU-birthrate}, and $\mathcal{B}_{\text{ETG}}$, Eq.~\ref{eq:ETG-birthrate}) with respect to  redshift, it is important to appropriately rescale the differential:
\begin{equation}
    \int_{t_0}^{t_f}\mathcal{B}(t)\diff t = \int_{z_0(t_0)}^{z_f(t_f)}\mathcal{B}(t(z))\left|\frac{\diff t}{\diff z}\right| \diff z \, ,
\end{equation}

\noindent where $\diff t/ \diff z = -1 / \left((1+z)H(z)\right)$. 
The integral of these birth functions 
($\mathcal{B} = \mathcal{B}_{\rm SFU}$ or $\mathcal{B} = \mathcal{B}_{\rm ETG}$)
is dimensionless and, being a normal distribution, it equals one. The ETG birth~function $\mathcal{B}_{\text{ETG}}$ in particular describes the unitary distribution of the birth of massive ETGs across redshifts per unit of cosmological volume. To derive the differential number of massive ETGs formed at a given redshift, it is necessary to account for the comoving volume. 
We employ comoving coordinates because on cosmological scales, the relative position of distant objects, such as galaxy clusters -- and, in our approximation as we will see in the next section, also of massive ETGs -- remains static in comoving coordinates. These objects move with the Hubble flow and maintain constant comoving positions. In this framework, the comoving separation between two such objects does not change over time, even as their proper distance $d_{\text{proper}}$ evolves due to the expansion of the universe. Namely, $d_{\text{proper}} = \frac{1}{1+z} d_C$, where $d_C$ is the comoving separation between two galaxies.

\subsubsection{Number density evolution of the ETGs}\label{sec:method-ETGcosmoVol-nz}
The present-day number density $n_0$ of massive ETGs is related to their average separation $\langle d_0 \rangle$ by:
\begin{equation}\label{eq:n0}
    n_0 = \frac{1}{\langle d_0 \rangle^3} \, ,
\end{equation}
which we saw in Sec.~\ref{sec:method-singleETG-ETGdistance} to be $\langle d_0 \rangle\approx 15$~Mpc. This distance averages, over cosmic volume, both the galaxy~cluster distance, as well as the average number of massive ETGs in a cluster. For any comoving volume at lower redshifts than the epoch of formation of massive ETGs, this number density is a constant. In the redshift range $20<z<15$, the number density will be the cumulative distribution function associated with the ETG~birth~function, rescaled by a factor of $n_0$. 
We note that we live in an under-dense region \citep{Keenan+2013}, and, as demonstrated by \citet{Haslbauer+2023}, the scale of cosmic inhomogeneity could extend to several Gpc. Furthermore, the authors found that overdensities at  $z=2$  could be 10 to 100 times higher than the local density \citep{Haslbauer+2023}. In such regions, the massive ETG separation could be significantly reduced, up to a factor of five (i.e., approximately $\langle d_0 \rangle\approx 3$~Mpc).

\subsubsection{ETG number differential}\label{sec:method-ETGcosmoVol-ETGnumber}

In comoving coordinates, the number density does not change with redshift, but the physical volume expands. However, the comoving volume the observer sees, which depends on the comoving distance, does increase as a function of redshift and for a given infinitesimal solid angle, ${\rm d}\Omega$, because an increasing number of regions of physical space are included with increasing redshift, and the light from more distant regions of space has the time to reach the observer. 
The comoving volume element, $dV_C(z)$, is given by \citep[e.g.,][their Eq.~28]{Hogg1999}:

\begin{equation}\label{eq:com_V_elem}
\diff V_C(z)\, =\, c \, \frac{D_C(z)^2}{H(z)} \, \diff \Omega \diff z\, ,
\end{equation}
where $H(z)$ is the Hubble parameter at redshift $z$, $\diff \Omega$ is the solid angle element in spherical~coordinates, $c$ is the speed of light, and $D_C(z)$ the comoving distance:
\begin{equation}\label{eq:com_dist}
D_C(z) = c \int_0^z \frac{\mathrm{d}z{\prime}}{H(z{\prime})}\,,
\end{equation}
which describes the distance to a light-emitting object as a function of $z$ in comoving coordinates.

The differential number of massive ETGs formed at a given redshift can finally be written as:
\begin{equation}
   \diff N_{\text{ETG}}(z) \, = \, \mathcal{B}_{\text{ETG}}\left(t(z)\right) \, \Big(n_0 \,  \diff V_C(z)\Big) \, ,
   \label{eq:ETGndiff} 
\end{equation}
which, when substituting in Eq.~\ref{eq:com_V_elem} and accounting for the full~sky, gives the rate of formation for massive ETGs:
\begin{equation}\label{eq:N_ETG_diff}
     \frac{\diff N_{\text{ETG}}(z)}{\diff z} \, = \, 4\pi \, c \, n_0 \, \mathcal{B}_{\text{ETG}}\left(t(z)\right) \,  \frac{D_C(z)^2}{H(z)}\,.
\end{equation}

The quantity $n_0 \diff V_c(z)$ remains the same in comoving and physical coordinates, given that the proper volume $\diff V_P$ scales as:
\begin{equation}
    \diff V_P = \frac{\diff V_c}{(1+z)^3} \, ,
\end{equation}

\noindent while the physical number density: 
\begin{equation}\label{eq:n0z}
    n_{\rm 0,P}(z) = n_0 \, (1+z)^3 \, ,
\end{equation}

\noindent if we let the separation $\langle d(z)\rangle = \langle d_0 \rangle \,(1+z)^{-1}$ evolve over time. The birth function $\mathcal{B}_{\text{ETG}}$ simply represents the distribution of $\diff N_{\text{ETG}}$ births as a function of time (or redshift). Therefore, it does not scale with redshift.

\subsubsection{Cumulative ETG luminosity evolution}\label{sec:method-ETGcosmoVol-ETGcumulative}
We can now obtain the cumulative bolometric luminosity of all the massive ETGs formed in the observable universe as a function of redshift, $\mathcal{L}_{\text{ETG, tot}}(t_{\text{cosmo}}(z))$, expressed as the convolution between the rate of formation of massive ETGs, $\diff N_{\text{ETG}}(z)/ \diff z$, with their average luminosity $\mathcal{L}_{\text{ETG}}(t)$:
\begin{align}\label{eq:Ltot}
    \mathcal{L}_{\text{ETG, tot}}(t_{\text{cosmo}}(z)) &= 4 \pi c n_0\int^{\infty}_{0} \bigg( \diff z'\frac{D_C(z')^2}{H(z')} \times \nonumber \\  &\mathcal{B}_{\text{ETG}}(t(z')) \left|\frac{\diff t'}{\diff z'}\right| \times \nonumber \\
    &\mathcal{L}_{\text{ETG}}
    \Big(t_{\text{cosmo}}(z) - t(z')\Big)\bigg)\,,
\end{align}
where the term $\left|\diff t/\diff z\right|$ has been renamed for consistency in the integrand.
The bottom panel of Fig.~\ref{fig:totETG} shows the evolution of $\mathcal{L}_{\text{ETG}}(t_{\text{cosmo}})$ as a linear function of cosmic time. The luminosity resulting from this equation holds for both comoving and physical coordinates, therefore it represents the total luminosity radiated by objects whose light will reach the observer at $z=0$. It can be treated as the intrinsic luminosity at time $t_{\text{cosmo}}$ (or redshift $z$).

\subsection{Energy density evolution of massive ETGs}\label{sec:method-Edens}

\subsubsection{The energy density evolution of the present-day CMB}\label{sec:method-Edens-CMBEdens}

For any black body, the total energy density at emission is related to Stefan-Boltzmann's law \citep[e.g.,][]{Rybicki1979}.
It can therefore be expressed as:
 \begin{equation}\label{eq:SBEdens}
     U_{\text{CMB},0} = \frac{4 \sigma}{c}T_{\text{CMB, obs}}^4 = 4.17 \times 10^{-14}
 \, [\text{J m}^{-3}] \, ,
 \end{equation}
 
 \noindent where $c$ is the speed of light, $\sigma$ is Stefan-Boltzmann's constant, and $T_{\text{CMB, obs}} = 2.7255$~K is the observed CMB temperature at present, i.e. $z=0$ \citep{Mather+1990, Planck2018}. 

In any FLRW-based cosmology, the temperature evolves with redshift $z$, and to retrieve the original temperature for a source emitting at redshift $z$, one applies $T_{\rm em}(z) = T_{\rm obs} (1 + z)$. In  standard cosmology, baryonic matter begins in a state of primordial plasma, primarily a mixture of H and He, with trace light elements like Li. As the universe expands, it cools. Hydrogen transitions into neutral gas at a temperature of roughly 3000~K. However, due to the high density in the early universe, photons can re-ionize hydrogen atoms, leading to a temperature at the surface of last scattering, from which the CMB originates, of $T_{\text{EoR}}\approx 2973$~K. This leads to an epoch of recombination redshift of $z_{\text{EoR}}\approx1090$. 
The energy density of the background we see today in the microwave range will therefore rescale as
 \begin{equation}\label{eq:SBEdens-z}
     U_{\text{CMB}}(z) = \frac{4 \sigma}{c}T_{\text{CMB, obs}}^4\, (1+z)^4 \, [\text{J m}^{-3}] \, .
 \end{equation}

\subsubsection{Photon energy density from the birth of massive ETGs at emission}\label{sec:method-Edens-ETGEdens}

Just as the CMB photon energy density can be quantified, so too can the photon energy density produced by elliptical galaxies. This is particularly relevant around the time of their formation, when their luminosity peaked.
We track a comoving volume -- e.g., 1~cMpc$^3$ -- where cMpc are comoving megaparsec units. 

In a first naive estimate, we can assume that all massive ETGs begin their evolution simultaneously.
The proper number density of massive ETGs in this comoving volume is described by Eq.~\ref{eq:n0z}.
The total energy each ETG radiates can be estimated by integrating Eq.~\ref{eq:LsingleETG} around the peak of star~formation:
\begin{equation}
    E_{\text{ETG}} = \int_{t_i}^{t_f}\mathcal{L}_{\text{ETG}}(t_{\text{ETG}})\, \diff t \, .
    \label{eq:integratedL}
\end{equation}

\begin{figure}[!tbp]
    \centering
\includegraphics[width=\columnwidth]{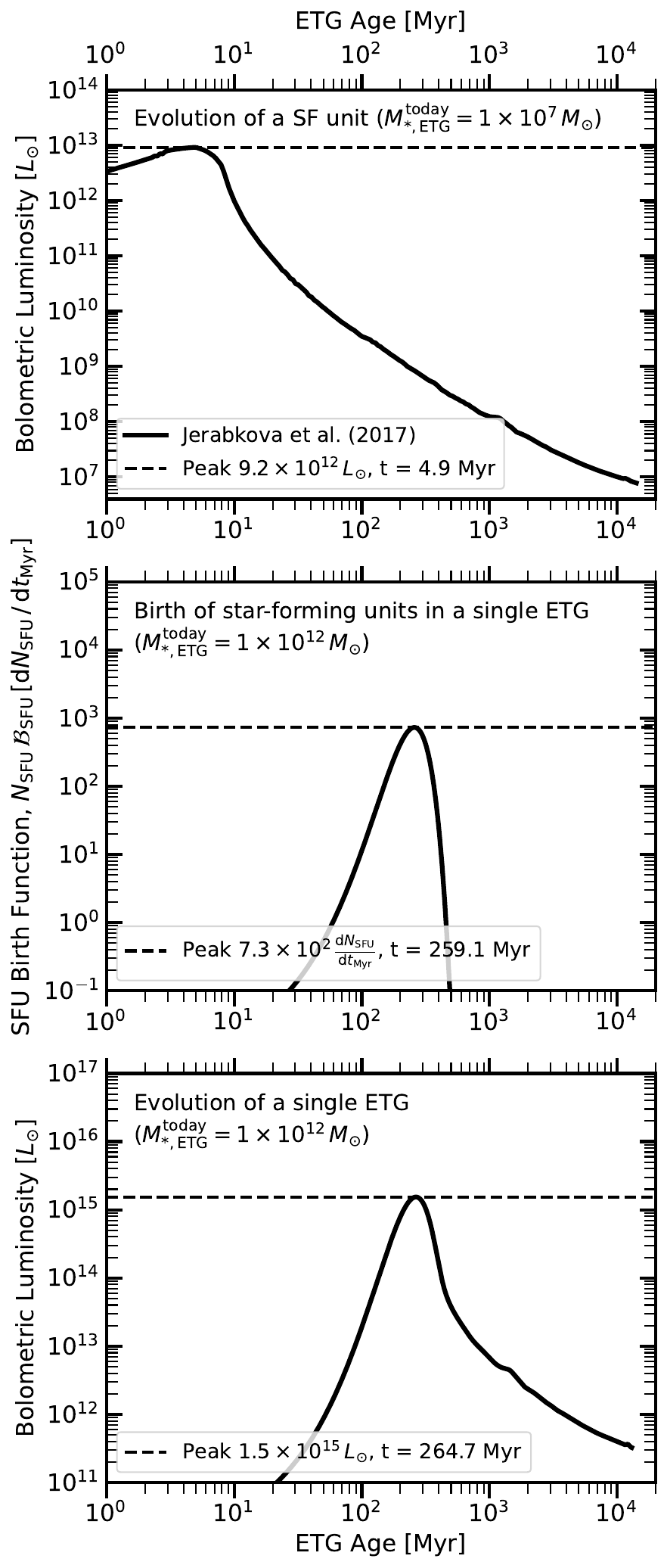}
    \caption{Evolution of a $10^{12}\, M_{\odot}$ massive early-type~galaxy (ETG), as described in Sec.~\ref{sec:method-singleETG}. The top~panel shows the bolometric luminosity evolution of a star-forming unit (SFU), the middle panel shows the stellar~birth~function, where the stellar mass is expressed in units of SFU. The bolometric luminosity of the whole ETG is shown on the bottom panel.}
    \label{fig:singleETG}
\end{figure}

\noindent $t_0$ and $t_f$ are extracted from the results of Fig.~\ref{fig:singleETG}, and they are identified as the times where $\mathcal{L}_{\text{ETG}}> 10^{13}\, L_{\odot}$. $\Delta t = t_f - t_i \approx 660$~Myr. Given the rapid evolution of the luminosity, and the short timescale when $\mathcal{L}_{\text{ETG}}$ is at its peak, the integrated luminosity (Eq.~\ref{eq:integratedL}) produces a similar total energy output throughout cosmic times.

In the naive case, the energy density at emission of massive ETGs within a proper volume of space can therefore be defined as:
\begin{equation}
    U_{\text{ETG}}(z) =  n_{0, \text{P}}(z) \, E_{\text{ETG}} \, ,
\label{eq:Uetg}
\end{equation}

\noindent where $n_{0, \text{P}}(z)$ was defined in Eq.~\ref{eq:n0z}. After converting solar luminosities to Watts, megayears to seconds, and megaparsecs to meters, can be expressed in SI units of Joules per meter cubed, $J \, m^{-3}$.

For the convolved luminosities, where massive ETGs form consecutively over a narrow interval rather than simultaneously (explained in Sec.~\ref{sec:method-ETGcosmoVol}), the total energy  density can be directly obtained by:
\begin{equation}
    U_{\text{ETG, conv}} = 
     n_0 \int_{t_a}^{t_b} \mathcal{L}_{\text{ETG, tot}} \diff t \, ,
\end{equation}

\noindent where $n_0$ comes from Eq.~\ref{eq:n0} and $\mathcal{L}_{\text{ETG, tot}}$ from Eq.~\ref{eq:Ltot}. $t_a$ and $t_b$ may span the age of the Universe, but we note that
the cumulative luminosity of all ETGs exceeds $\mathcal{L}_{\text{ETG, conv}}> 10^{20} L_{\odot}$ (i.e., the lower panel of Fig.~\ref{fig:totETG}), in a range where $\Delta t = t_b - t_a \approx 900$~Myr.

\begin{figure}[tbp]
    \centering
    \includegraphics[width=\columnwidth]{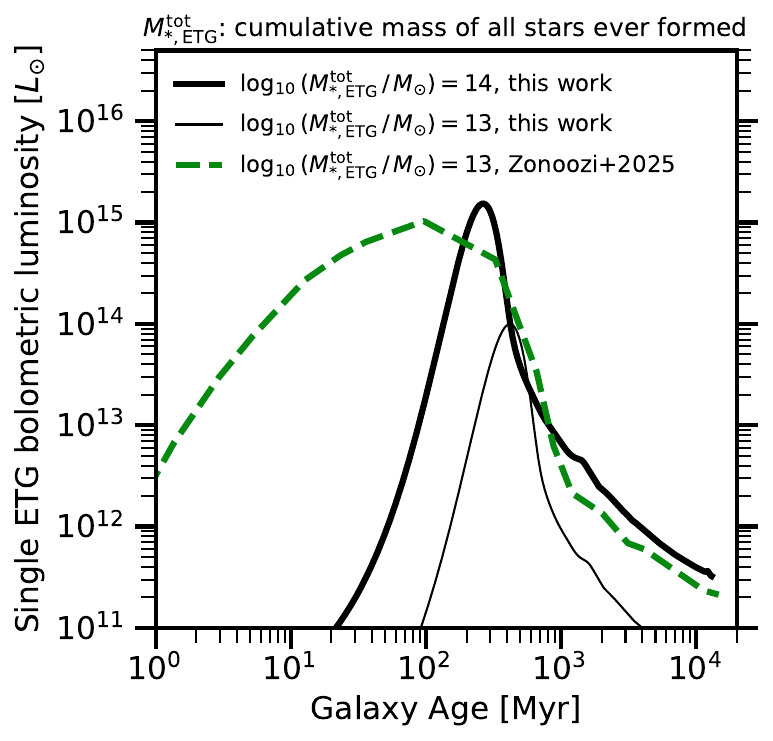}
    \caption{Comparison of the bolometric luminosity evolution of massive ETGs in this work (\emph{solid black curves}) with published theoretical models (\emph{dashed green line}). The thicker solid line is computed assuming a total cumulative stellar mass (Eq.~\ref{eq:totcumstellarmass}, including the mass of all stars ever formed -- both surviving and long-dead), of $M_{*, \text{ETG}}^{\rm tot} = 10^{14}\, M_{\odot}$, and is the same curve shown in the bottom panel of Fig.~\ref{fig:singleETG}. The thinner solid line represents the bolometric luminosity evolution obtained using the same method but for a total cumulative stellar mass of $M_{*, \text{ETG}}^{\rm tot} = 10^{13} \, M_{\odot}$.
    For comparison, we include results from \citet{Zonoozi+2025} (their Fig.~9), who employ an empirically driven variable galaxy-wide~IMF model based on the IGIMF theory \citep{KroupaWeidner2003, Kroupa+2013} to compute the bolometric luminosity evolution of a galaxy with a total cumulative stellar mass of $M_{*, \text{ETG}}^{\rm tot} = 10^{13} \, M_{\odot}$. 
    }
    \label{fig:Zoonozi}
\end{figure}

\subsubsection{Thermal equilibrium of the ETG luminosity with cosmic dust emission}\label{sec:method-Edens-ETGdust}

Star~formation is accompanied by nearly instantaneous dust~formation on cosmic timescales. As discussed in Sec.~\ref{sec:intro}, the lifetimes of massive stars are shorter than those of the molecular clouds hosting the embedded clusters where such stars are born. Dust~formation occurs on even shorter timescales. The well-studied case of SN~1987~A, whose progenitor was a 20~$M_{\odot}$ blue supergiant star, produced 0.6-8~$M_{\odot}$ of dust within decades of its detection \citep[e.g.,][]{Wesson+2015}. Similarly, SN~2010jl formed comparable dust masses within just a year of the explosion \citep{Gall+2014}. 

In the Early Universe, stellar~nucleosynthesis acts as the primary bottleneck to dust~formation \citep{CherchneffDwek2009}, because the dense shockwave environments of supernovae rapidly saturate refractory species \citep{CherchneffSarangi2017}. At the same time, intense SFR such as the ones that form massive ETGs are naturally enriched rapidly with chemical species \citep{Yan+2021}, because the production of metals is the byproduct of thermonuclear processes as well as stellar feedback. Stellar mergers in the extremely dense and massive embedded clusters (our SFUs) eject a significant amount of processed material -- even prior to any core-collapse supernovae -- through the energetic binary-binary encounters of massive stars. This is a primary reason for the observed multiple populations in present-day globular star clusters (see \citealt{Wang+2020}). 
As a consequence of these intense stellar-dynamical processes, chemical enrichment as well as dust formation are expected to occur with the onset of the formation of massive stars in the SFUs. The ``\emph{missing quasar problem}'' may indeed be evidence that the  early star-forming Universe was heavily enshrouded in optically thick cosmic dust \cite{Kroupa+2020}. And indeed, recent observations from the A3COSMOS database constrained in the redshift range $6 < z < 12$ show that a significant portion of the dust production  has already occurred by $z=7$ \citep{Ciesla+2024}. Some of the dust will be driven to larger distances from the star bursts through radiation pressure \citep[e.g.,][] {NakazatoFerrara2024}.

It is therefore reasonable to assume that dust was sufficiently abundant to thermalize the extreme radiation permeating the medium during these early epochs. However, we note that intense stellar~feedback is accompanied by rapid dust~destruction \citep[e.g.,][]{Draine1995}, leaving no direct evidence of the existence of this dust at present. The only indicators  
of this process are the supersolar metallicities observed in massive ETGs, and the gray-body radiation that this early dust is expected to have emitted.

A word must be said about the typical dust temperatures encountered in various astrophysical environments \citep{Ferriere2001, Draine2011}. Dust survival can be evaluated using temperature-pressure phase diagrams \citep{Whittet2003}, which show that dust can survive in stability zones at temperatures up to a few 1000~K. Temperatures of about a thousand kelvin are indeed observed in proto-planetary disks \citep[e.g.,][]{Abraham+2009}.
In star-forming main-sequence galaxies, predominantly spiral galaxies, the typical dust temperature is around 20~K \citep{Draine1990, Planck2016dust}, which is commonly adopted in spectral energy distribution modeling tools. In contrast, in the dense wave-fronts of H~II supernova bubbles, corresponding to photo-dissociation regions, dust temperatures range from 19 to 26K \citep{Anderson+2012PDRdust}. However, \citet{Contini+2003} have shown that in star-forming regions, the temperature of thermalized dust can fluctuate significantly, reaching often temperatures of 300~K.
While laboratory experiments offer stringent constraints into dust properties, the actual dust temperature in astrophysical environments spans as wide a range as the diverse locations where dust is found. To take average values, it can be said that the warm dust component in starburst regions typically spans temperatures from 40 to 55~K \citep{Calzetti+2000}.

A full analysis of the radiation field at the epoch of the massive ETG formation would involve solving the radiative transfer equations \citep[e.g.,][]{Rybicki1979}, accounting for gas and dust absorption coefficients, as well as dust emissivity \citep[see][for a detailed analysis]{Zhang+2016}. Here, however, we consider the simplest scenario in which the bolometric luminosity of the massive ETG is fully thermalized by dust, with no energy loss ($L_{\text{bol}} \approx L_{\text{dust}}$). Given that the flux is a luminosity over the surface area it crosses -- i.e., the surface of a sphere around the galaxy -- it follows that for any massive ETG:
\begin{equation}
    L_{\text{dust, ETG, em}} = 4 \pi R^2_{\text{eff}}\sigma T_{\text{dust, ETG, em}}^4\,,
\end{equation}
where $R_{\text{eff}}$ is the effective radius of the dust-emitting region, and $T_{\text{dust, ETG, em}}$ is the temperature of the thermalized dust in the massive ETG at emission. Even for a massive ETG, $R_{\text{eff}}$ should be tightly bound around the galaxy. If we consider the peak luminosity of a massive ETG ($\approx 10^{15}\, L_{\odot}$, Fig.~\ref{fig:singleETG}) whose effective radius is $R_{\text{eff}}\approx 10 \, \text{kpc}$, $T_{\text{dust, ETG, em}} \approx 50 \, \text{K}$, which not only is comparable to the CMB temperature at redshift 17, $T_{\text{CMB}}(z\approx 17)$, but it is also comparable to the typical dust temperatures observed in starburst galaxies.

We re-emphasize at this point that the objective of this study is to reconstruct, using insights from the literature on galaxy and stellar evolution, the potential contribution of the formation of massive ETGs to CMB foreground subtraction. The proximity of the values here estimated emerge naturally from the theory, without any CMB priors or constraints. Details such as the precise thermalization temperature of the dust and the impact on CMB anisotropies are left for future works.

\section{Results\label{sec:results}}

\subsection{The evolution of the bolometric luminosity  of massive ETGs and of their ensemble}\label{sec:results-ETGs}

\begin{figure}[!tbp]
    \centering
    \includegraphics[width=\columnwidth]{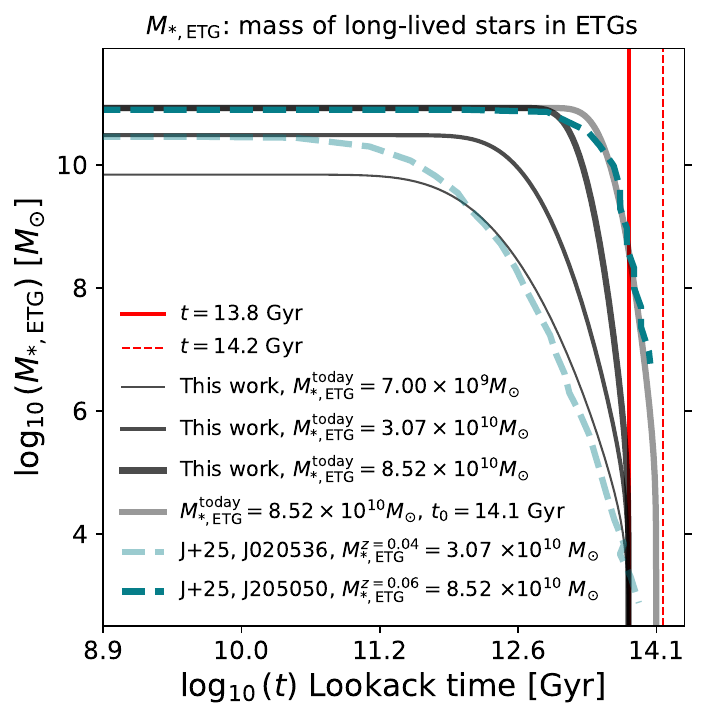}
    \caption{
    Evolution of the stellar mass of individual ETGs in this work (\emph{solid black and gray lines}) compared to observational inferences (\emph{dashed teal lines}, \citealt{Jegatheesan+2025}). Notably, \citet{Jegatheesan+2025} report galaxies that appear to have started forming over $14$~Gyr ago, earlier than the $13.8$~Gyr age of the Universe inferred by the Planck Collaboration. Reference ages are marked with \emph{red vertical lines}, as labeled. The \emph{thick lines} from this work  both assume the same present-day stellar mass, $M_{*,\text{ETG}}^{\rm today}$, same as in the galaxy ``J205050" from \citet{Jegatheesan+2025}. Both thick lines are identical in shape but assume an origin either  at $t_0 = 13.8$~Gyr or at $14.1$~Gyr. The \emph{thin lines}, which match the $M_{*,\text{ETG}}^{\rm today}$ of ``J020536", both assume an origin at $t_0 = 13.8$~Gyr.}
    \label{fig:Jegatheesan}
\end{figure}

The results from Sec.~\ref{sec:method-singleETG} are encapsulated in Fig.~\ref{fig:singleETG}. The top panel shows the bolometric evolution of a SFU, as was derived by \citet{Jerabkova+2017}. The present-day stellar mass of a SFU is about $M_{*,{\rm SFU}}^{\rm today}= 10^7 M_{\odot}$. However, as exemplified by the IGIMF \citep{KroupaWeidner2003, Jerabkova+2018, Kroupa+2024}, the SFU will have produced a total of $M_{*,{\rm SFU}}^{\rm tot}=10^9 M_{\odot}$ in stars throughout its lifetime. Notice that the SFU evolves very rapidly, reaching its peak luminosity only after $\approx 5$~Myr. The SFU is characterized by a brief and intense ($> 10^2\,M_{\odot}{\text{/yr}}$) SFR. The peak bolometric luminosity of this dwarf compact object nearly reaches $10^{13}\, L_{\odot}$, highlighting the striking range of bolometric luminosities caused by the SFR in a SFU during its evolution, spanning 6 orders of magnitude. We stress that massive globular clusters or ultra-compact dwarf galaxies must have attained such luminosities in order to have formed the mass in stellar mass black holes as inferred from their present-day dynamical mass-to-light ratios \cite{Dabringhausen+2009, Jerabkova+2017, Mahani+2021}.

The middle panel of Fig.~\ref{fig:singleETG} shows the birth function of the SFUs in a single average massive ETG, assuming that the masses of all the SFUs are uniformly distributed, and that they are born with a Gaussian distribution, as was explained in Sec.~\ref{sec:singleETG-birthfunction}. This birth function was calculated for a massive ETG whose final stellar mass is roughly $M_{*,ETG}^{\rm today}= 10^{12} M_{\odot}$. At the star~formation peak, this galaxy will form a few hundred SFU per Myr.

Convolving the top panel with the middle panel returns the total bolometric luminosity for this average massive ETG, shown in the bottom panel of Fig.~\ref{fig:singleETG}. As expected due to the rapid luminosity evolution of a single SFU, the peak luminosity is reached shortly after the peak in SFR. And in this case, just as for a single SFU and consistent with \citet{Jerabkova+2017}, we find that the whole massive ETG will span a few ($\approx4$, in this case) orders of magnitude between the peak and present-day luminosity throughout its lifetime.

Similar results are shown in Fig.~\ref{fig:totETG}, where the top panel represents the birth function of all massive ETGs, assuming that any massive ETG ($M_{*,\text{ETG}}^{\rm today}> 10^{11.5}M_{\odot}$) formed in the redshift range $15<z<20$, roughly coincident with the 21-cm anomaly. 
Most interestingly, we see that the birth function informs us that more than 20 million ETGs will be forming per Myr at the peak of the ETG formation. 
The convolved luminosity evolution of all massive ETGs is shown in the bottom panel of the same figure. We see that the peak luminosity appears with a delay that roughly encompasses the time it takes for an ETG to form, as well as the formation timescale for the birth function of all ETGs.

As was explained both in Sec.~\ref{sec:intro} and Sec.~\ref{sec:method-singleETG-LumEvol}, $\Lambda$CDM with its predictions of structure formation and time-to-redshift conversion is unlikely to remain viable for much longer.
We do not expect these short timescales to remain as tight in the future generation of cosmological models.

To assess the reliability of our approximation for estimating the luminosity of individual ETG galaxies, we compare our single ETG bolometric evolution with the fully detailed luminosity evolutions derived from the IGIMF theory \citep{Zonoozi+2025} in Fig.~\ref{fig:Zoonozi}. This comparison demonstrates that our method provides a conservative estimate of the bolometric luminosity evolution of massive ETGs, as more detailed models incorporating extensive observational constraints yield bolometric luminosities up to an order of magnitude higher for the same total stellar mass,  $M_{*, \text{ETG}}^{\rm tot}$. 

In Fig.~\ref{fig:Jegatheesan}, we compare the evolution of long-lived stars in single massive ETGs from our model with two massive ETGs analyzed by \citet{Jegatheesan+2025}. These authors used MUSE integral field unit (IFU) spectra to construct 2D stellar population maps for ETG targets. In an earlier study, \citet{Jegatheesan+2024} found clear observational evidence of downsizing in the bulge components of nearly 1000 spiral galaxies. Expanding on this, \citet{Jegatheesan+2025} examined ETGs, reporting the flux ratio between the inner and outer components. We focus on the inner component, which follows an evolutionary path consistent with monolithic collapse. For illustrative purposes, we select two galaxies, J020536 ($z = 0.04$) and J205050 ($z = 0.06$), using their total observed stellar masses at their respective redshifts.\citet{Jegatheesan+2025} reports flux ratios between the inner and outer components (their Table~3). We assume both components share the same $M/L$ ratio due to the old ages of both components. We then reconstruct their stellar mass evolution using the cumulative mass fractions  from their Figs.~7~and~8, multiplied by the rescaled stellar masses (from their Table~1). 
The observed elliptical galaxy J020536 has a smaller inner component ($34$\% of the total stellar mass) compared to J205050, whose inner component constitutes $45$\% of the total. 
Consequently, J020536 exhibits a more extended evolution timescale, possibly reflecting contamination from the formation of the outer component. In contrast, J205050 follows a path remarkably consistent with our model and downsizing timescales, making it a good candidate for monolithic collapse.
In our model, solid black lines vary in thickness to indicate different final masses. The both thick lines correspond to  $M_{*,\rm ETG}^{\rm today} = 8.52 \times 10^{10} M_{\odot} $ but while the black one begins its evolution at 13.8~Gyr, the gray curve is shifted to begin at an earlier epoch ($14.1$~Gyr) to match the onset of J205050.
For our model curves, we integrate the rate of SFU formation (middle panel of Fig.~\ref{fig:singleETG}), yielding  $N_{\rm SFU}(t)$, which we then weight by the mass of long-lived stars per SFU to obtain the mass evolution of long-lived stars in each ETG. Our model best describes the most massive ETGs, which likely experienced high SFRs sufficient to form ultra-compact SFUs. In lower-mass ETGs, where SFRs are only a few  $M_{\odot}/\mathrm{yr}$, the galaxy-wide IMF is expected to resemble the canonical form of the stellar IMF \citep{Kroupa2001}, leading to a higher $M/L$ ratio during star-forming epochs compared to more massive ETGs.
\citet{Kroupa+2020} (their Fig. 10) illustrate how  $M_{*, \rm ETG}^{\rm tot}$  (our Eq.~\ref{eq:totcumstellarmass}, their  $M_{\rm igal}$) correlates with  $M_{*, \rm ETG}^{\rm today}$ (our Sec.~\ref{sec:method-singleETG-ETGproperties}, their  $M_{\rm pgal}$).

\subsubsection{Number of massive ETGs per Planck pixel}

\begin{figure}[!tbp]
    \centering
\includegraphics[width=\columnwidth]{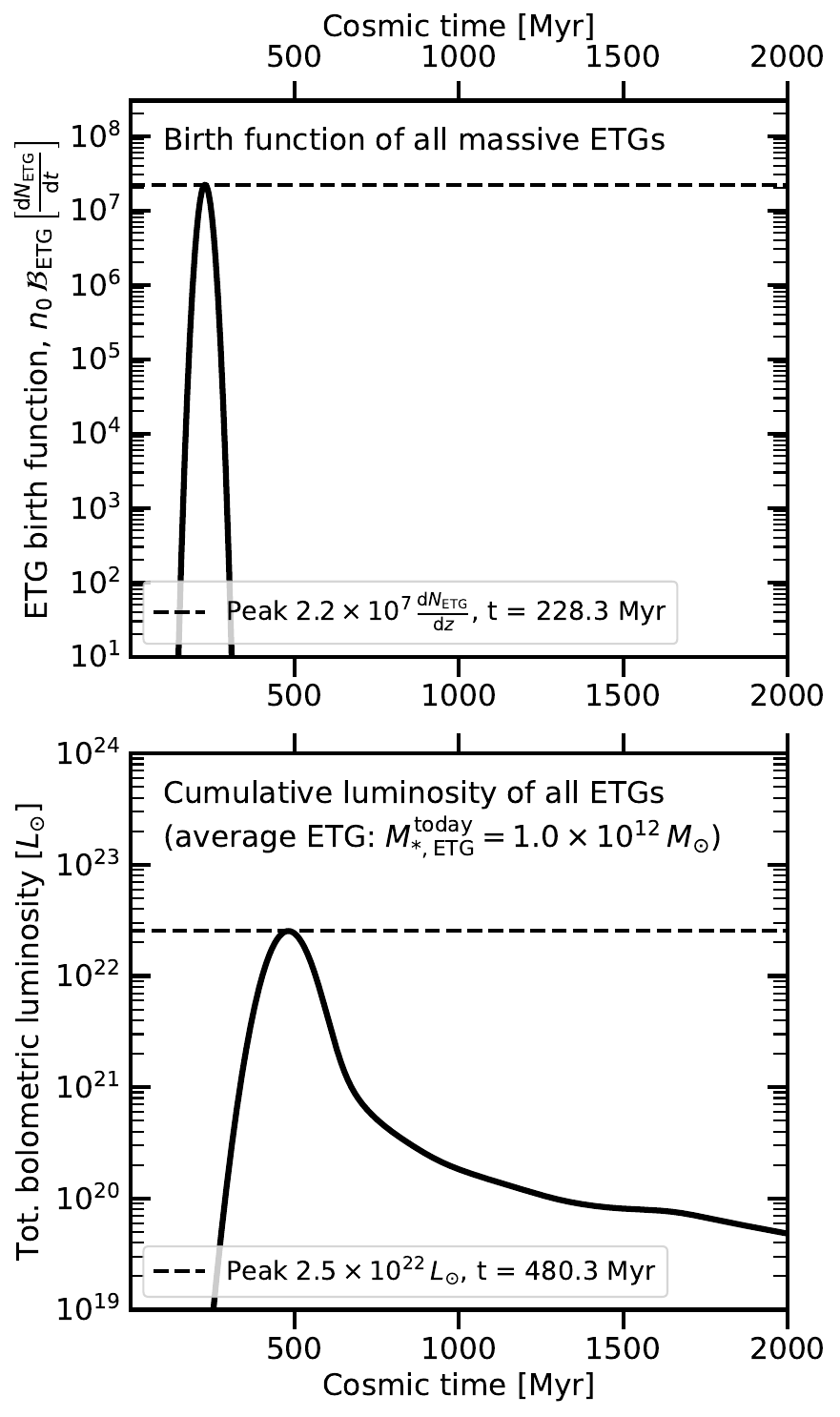}
    \caption{Evolution of all massive ETGs, as described in Sec.~\ref{sec:method-ETGcosmoVol}. The \emph{top panel} shows the ETG birth function, constrained so that $\pm 3 \sigma$ of all present-day massive ETGs were born in the redshift range $15 < z < 20$. The \emph{bottom panel} shows the total convolved bolometric luminosities of all massive ETGs (Eq.~\ref{eq:Ltot}). In both plots the quantities are shown to evolve with respect to linear cosmic time.}
    \label{fig:totETG}
\end{figure}

The birth~function shown on the top panel of Fig.~\ref{fig:totETG} spikes and evolves quite rapidly. Its integral represents the total number of galaxies whose luminosity contributes to the spike then depicted in the lower panel. The integral of this figure returns $N_{\text{ETG}} \approx 4.4 \times 10^7$~massive~ETGs, distributed over the full sky, i.e. $4 \pi$~steradians. The resolution of the Planck satellite is between 5 to 10 arc-minutes, depending on the frequency band \citep{Planck2005}. Given that, for the whole sky, its total area is:
\begin{align}
    A_{\text{sky}}& = 4 \pi \, \text{[steradians]} \, \times \left(\frac{180}{\pi}\right)^2  =  4.125 \times 10^4 \, \text{deg}^2  \nonumber \\ 
     & = 4.125 \times 10^4 \, \text{deg}^2 \, \times \frac{3600 \text{arcmin}^2}{\text{deg}^2}\nonumber \\
     & =1.49 \times 10^8 \, \text{arcmin}^2 \, ,
\end{align}

\noindent we can estimate how many Planck resolution elements are contained in the full sky. In the case of Planck's best resolution, 5~arcmin, we get an area of:
\begin{equation}
    A_{\text{Planck}} = \left(\frac{5'}{2}\right)^2 \pi \approx 19.6 \, \text{arcmin}^2 \,,
\end{equation}

\noindent leading to a number of resolution elements $N_{\text{Planck, res}} = 7.58 \times 10^6$. Assuming that massive ETGs are distributed isotropically, each Planck pixel will contain about 6 massive ETGs whose luminosity contributes to the peak in Fig.~\ref{fig:totETG}.

\subsection{Energy density from forming ETGs}\label{sec:results-Edens}

\subsubsection{The intrinsic energy density from forming ETGs}\label{sec:results-presentdayEdens}

In  Fig.~\ref{fig:CMBEdens} we present the main results of this work. It depicts energy densities both at emission (the markers) and as they evolve with redshift. The energy density of the primordial plasma at the epoch of recombination was 0.06~J~m$^{-3}$, and it evolved to the present-day $4.18 \times 10^{-14}$~J~m$^{-3}$.

{\renewcommand{\arraystretch}{1.5}
\begin{table}[ht]
    \centering
    \begin{tabular}{c c c c }
    \hline
        $\langle d_0 \rangle$ [Mpc] & 9  & 10  & 15   \\ \hline 
        $U_{\text{ETG}}(z=0)$ [$\times 10^{-16}$ J m$^{-3}$] & $\, 1600\, $ & $\, 630\, $ & $\, 16\, $ \\  
        $U_{\text{ETG, conv}}(z=0)$ [$\times 10^{-16}$ J m$^{-3}$] & $\, 580 \, $ &$\, 230 \, $ & $\, 5.9 \, $\\ \hline
    \end{tabular}
    \caption{
    Present-day energy densities ($U_{\text{ETG}}(z=0)$, Eq.~\ref{eq:Uetg}) computed according to Sec.~\ref{sec:method-Edens-ETGEdens}, assuming different present-day massive ETG separations, $\langle d_0 \rangle$. \citet{Haslbauer+2023} found large inhomogeneities at scales of a few Gpc, where the present-day average separation of massive ETGs could be as small as 9~Mpc. These estimates are to be compared to the present-day energy density of the CMB: $U_{\text{CMB}}(z=0)=418 \times 10^{-16}$~J~m$^{-3}$.
     }
    \label{tab:U}
\end{table}
}

\subsubsection{A consistency check: young massive ETGs resemble a black body}\label{sec:results-consistency}

\begin{figure*}[!tbp]
    \centering
\includegraphics[width=0.8\textwidth]{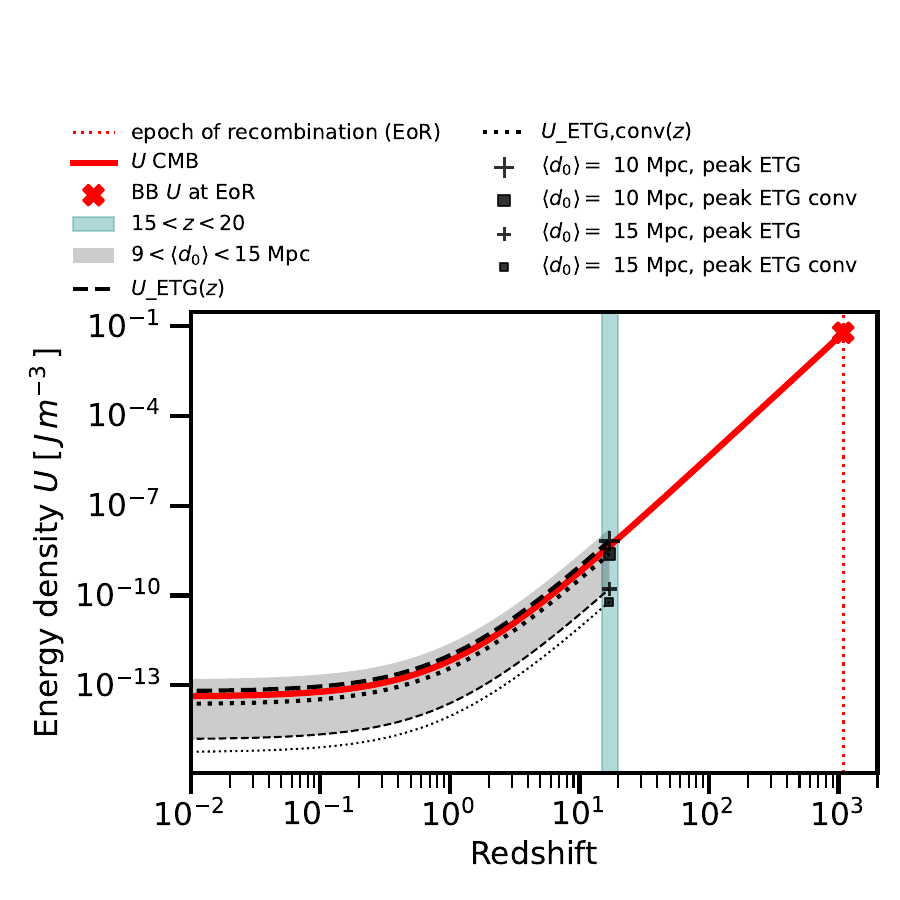}
    \caption{This figure represents the evolution (\emph{curves}), as a function of redshift of the energy densities ($U_{i,\text{em}}$) at emission (\emph{markers}) of multiple sources $i$, namely primordial plasma, massive ETGs born simultaneously, and cumulative massive ETGs. The \emph{red bold cross} represents the energy density of a uniform and isotropic blackbody, observed today with a temperature of 2.7255~K, rescaled by $(1+z)^4$ to its value at the surface~of~last~scattering, $z_{\text{EoR}}=1090$. The \emph{black plus signs} (\emph{peak ETG}) represent the energy density at emission under the naive approach, i.e., all the massive ETGs were born simultaneously. The \emph{black squares} (\emph{peak ETG conv}) are the peak of the cumulative energy density of all ETGs as a function of redshift when the luminosity of massive ETGs is convolved with the ETG birth function. The \emph{smaller markers} refer to an average massive ETG separation $\langle d_0 \rangle =$~15~Mpc, while the \emph{larger markers} refer to $\langle d_0 \rangle =$~10~Mpc. The \emph{vertical shaded region in teal}  corresponds to the $15<z<20$ redshift range. The \emph{gray shaded region} spans the full range of $\langle d_0 \rangle$ in the naive approach. The largest separation is that of the local value ($\langle d_0\rangle \approx 15$~Mpc). At $z\approx 2$ the separation can be as small as  $\langle d_0\rangle \approx 9$~Mpc \citep{Haslbauer+2023} as explained in Sec.~\ref{sec:method-Edens-ETGEdens}). All these  energy densities evolve from emission according to $U_i(z)=U_{i,\text{em}}(1+z)^{-4}$. The most conservative estimate, namely $\langle d_0 \rangle = 15$~Mpc in the convolved case, evolves to be 1.4\% of the present-day energy density of the CMB (fine black dotted line vs red line). If instead we assume the average separation at $z \approx 2$, consistent with \citet{Haslbauer+2023}, the energy density from elliptical galaxies matches that of the CMB (thick dashed and dotted black lines vs red line). 
    }
    \label{fig:CMBEdens}
\end{figure*}

Given the extreme brightness of massive ETGs during their formation and their relative proximity at that epoch, it becomes unavoidable to consider their contribution to the background radiation. This radiation estimated in Sec.~\ref{sec:method-Edens-CMBEdens}, redshifted over time, is now observed in the microwave range. Here we perform a consistency check of this calculation.
The CMB energy density $U_{\text{CMB}}$ was calculated in Eq.~\ref{eq:SBEdens}. 
Along with the results obtained in terms of the 
energy density
(Fig.~\ref{fig:CMBEdens}), another estimate can therefore be obtained concerning the possible contribution of massive ETGs to the present-day observed CMB photon energy density. 
As we have seen, dust destruction timescales are extremely short, and comparable to the lifetime of a few $10^7$~yr. The timescales shorten significantly in the presence of strong irradiation \citep{Slavin+2020}. On cosmic timescales, it is reasonable to assume that dust is destroyed shortly after reaching thermal equilibrium.
In each co-moving Gpc$^3$ volume there are $N_0 = n_0 \times \, \text{ 1 Gpc}^3 \,\approx 3.3 \times 10^5\,$ETGs emitting at their dust temperatures of $40< T_{\rm dust, ETG, em}\, [\text{K}]<55$ (Sec.~\ref{sec:method-Edens-ETGdust}). This range of dust temperatures is consistent with the average dust temperature in a massive ETG, $T_{\text{dust, ETG, em}}\approx 50$~K, adopting a fiducial effective radius of about 10~kpc around the ETG's luminosity peak. This range is also reported from high-redshift observations of star-bursts at high redshift \citep[see the discussion and references therein]{Kroupa+2020}.

At $z=17$, the emitting sources can thus be approximated as being a continuum. The emission of photons by the ETGs therefore becomes the same problem in terms of their contributing energy density as in the case of the CMB which was also emitted near-instantly at $z_{\text{EoR}} \approx 1090$. Therefore, ETGs contribute an energy density of dust-thermalized photons at emission near $z=17$ of:
 \begin{equation}
     U_{\text{ETG, em}} = \frac{4 \sigma}{c}T_{\text{dust, ETG, em}}^4 \,,
 \end{equation}
 where $T_{\text{dust, ETG, em}}(z)= (1+z)\, T_{\text{dust, ETG, obs}}$ with $z=17$. For average ETG peak luminosities of $\approx 10^{15} L_{\odot}$, $T_{\text{dust, ETG, obs}} \approx 2.7$~K. It follows that $U_{\text{ETG},0} \approx U_{\text{CMB},0}$ verifying the more accurate calculation shown in Fig.~\ref{fig:CMBEdens} that ETGs appear to have provided a flux in photons that is naturally in agreement with that observed in the CMB.

Massive ETGs exist and can be measured locally. We know they have formed rapidly at high redshift, and we now find them to contribute a comparable energy density as what is observed in the CMB photon field today.

\section{Discussion\label{sec:discussion}}

The greatest achievement of precision cosmology has been the measurement of the anisotropies observed in the cosmic microwave background (CMB). In particular, it is possible to obtain a well-constrained multi-parameter fit for $\Lambda$CDM as well as its popular more elaborate variants, particularly so with the advent of the technical triumphs of the Planck Collaboration \citep{Durrer2015}.
Those CMB anisotropies are understood to be the seeds that later evolved into large-scale structures under the influence of gravity. They are believed to have originated from the primordial quantum fluctuations within a Hot Big Bang framework \footnote{In this framework, the Universe began as a hot plasma with all of its matter contained in a dense original volume, and the CMB is the radiation liberated from the transition of this plasma to a neutral state.}. However, the fluctuations would have occurred at subatomic scales, with the Planck length ($10^{-35}$ m) being the typical assumed scale. Meanwhile, the average sound horizon at the time of recombination was 138 kpc \footnote{The sound horizon is the maximum distance that acoustic waves could have traveled in the early universe's plasma before the plasma cooled to a neutral state. It represents therefore the average physical separation between the regions corresponding to the first peak of the CMB anisotropies.}. To explain the stark size difference between the scale of the CMB anisotropies and the primordial fluctuations, it is therefore imperative to pair $\Lambda$CDM with inflation \citep{Guth1981}, i.e., the exponential adiabatic expansion that must have occurred within $10^{-32}$~seconds after the Big Bang \citep[for a review, see ][]{Turner2022}. Within this brief instant, inflation would swiftly expand the scale factor of the universe by at least 60 \emph{e}-folds, i.e. a volume increase by a factor of $10^{78}$ \citep{LiddleLyth2000}. 
In comparison, the volume increase driven by $\Lambda$CDM from 
the electroweak era ($10^{15}$~K) to recombination ($\approx3000$~K) is a factor of $10^{34}$ over $10^{12}$~s, while from 
recombination to the present, the volume increases by 
a factor of $10^9$ over $4\times10^{17}$~seconds. 
Inflation is the crucial link that allows interpreting the CMB as the relic of a Hot Big Bang. 
However, inflationary models are strained by observational constraints \citep[for a recent review, see][ where 40\% of models have been excluded by data
]{Martin+2024}. Also notable is that the angular two-point correlation function observed by the Planck collaboration \citep{Planck2016} remains surprisingly close to zero for angles larger than 60 degrees \cite{Schwarz+2016}, contrary to theory which expects also these regions to be causally connected. 

The full CMB signal is heavily contaminated by foreground Galactic and extragalactic sources. Data cleaning and analysis inevitably depend on the assumed cosmological and astrophysical models, introducing a bias when using the CMB data to test the input cosmology. The Planck Satellite Scientific Programme \citep{Planck2005} stressed the risk for over-cleaning, which can lead to mistakenly treat genuine sources as contaminants. Conversely, these contaminating ``\emph{massive objects are anomalous and could cause gravitational lensing of the surface of last scattering in excess of the standard calculation made in CMB fits}'' \cite{McGaugh2024b}.

\section{Conclusions\label{sec:conclusions}}

We assume the standard cosmological $\Lambda$CDM model, which stringently requires the CMB to be the photons emitted from the epoch of reionization at $z_{\text{EoR}} \approx 1090$.  By incorporating the knowledge gained over the past decades on the properties and evolution of massive ETGs, it transpires that these  objects formed at $z>9$, within a few hundred Myr and with galaxy-wide IMFs highly dominated by massive stars. These constraints are requirements in order to attain the super-Solar metallicities observed in massive ETGs. 

These conditions imply that massive ETGs were some-$10^4$ times brighter than their present-day luminosity.  Their fast and intense primordial star formation drove early chemical enrichment and the rapid production of dust. The dust would subsequently reach the thermal equilibrium with the intense radiation field. Here we calculate the energy density of that thermalized photon field. 
Under a monolithic framework, we find that
each massive ETG should form from the gravitational collapse of a cloud whose radius at the time of formation was about 400 kpc (Sec.~\ref{sec:method-singleETG-ETGdistance}). If the average separation of massive ETGs is the local value of about $\langle d_0 \rangle \approx 15$~Mpc, then their epoch of formation would be at a redshift of $z\approx 17$. 
Today, the radiation produced by massive ETGs in this scenario would produce a photon field approaching that of the observed CMB photon energy density. 

First, we present a reasonable analytic evolution model for the bolometric luminosity of massive ETGs (Sec.~\ref{sec:method-singleETG}).
Then, we add up the bolometric luminosity of all massive ETGs in a cosmological volume (Sec.~\ref{sec:method-ETGcosmoVol}). 
We consider two scenarios, one is the ``naive'' approach, where all massive ETGs are born simultaneously. In the other ``convolved'' scenario (Sec.~\ref{sec:method-ETGcosmoVol-ETGcumulative}), we propose a short early epoch of formation of all ETGs. We then provide rough first-order estimates of the energy released by this intense starburst scenario (Sec.~\ref{sec:method-Edens}). 

What results from these estimates, naturally but unexpectedly,
is as follows. Under the most conservative assumptions -- i.e., assuming that the average ETG separation ($\langle d_0 \rangle$) in local volumes applies to the whole Universe -- the formation of massive ETGs may account for 1.4\% of the observed (present-day) CMB photon energy density. 
If, instead,
$\langle d_0 \rangle$ corresponds to that inferred around $z\approx 2$  \citep{Haslbauer+2023}, then the cumulative energy density of massive ETGs would be of the same order as that of the observed CMB (Fig.~\ref{fig:CMBEdens}).
Assuming that dust produced by massive ETGs reaches thermal equilibrium with the ETG radiation field, the resulting emission would form a background near $z \approx 15$ that, when redshifted, appears to observers at $z=0$ as a microwave photon field (Sec.~\ref{sec:results-Edens}).
The independent data now coming from the observations with the JWST of the formation of massive galaxies at $z>9$ and the ALMA observations of high-redshift, dust-rich, star-forming galaxies support this conclusion, as do the recent observationally obtained mass-growth times and rates of individual elliptical galaxies (Fig.~\ref{fig:Jegatheesan}).

The results obtained for massive elliptical galaxies suggest they are merely embers in the ashes of ancient cosmic bonfires.
i.e., their mass consists mostly of stellar-mass black holes, neutron stars and old white dwarfs. Their present-day luminous main~sequence stars constitute merely some 0.1--1~per cent of the total stellar mass formed in them and only about 10--20~per cent of their present-day baryonic mass (Fig.~\ref{fig:singleETG}, bottom panel). This is a direct consequence of the galaxy-wide IMF, which needs to be dominated by massive stars in order to ensure the nucleosynthesis of the large chemical abundances observed in the stellar spectra of these galaxies. The dominance of stellar remnants is evident in the approximately tenfold~larger V-band dynamical mass-to-light ratios of massive ETGs  \citep[e.g.,][their Fig.~10]{Yan+2021}.

Even if the absorption coefficients and dust emissivity were not highly efficient, should only a few percent of the peak luminosity from massive ETGs make its way to us, it would have a profound impact on the interpretation of the CMB power spectrum. Nonetheless, dust is expected to process radiation very efficiently. In starburst galaxies, a comparison between infrared luminosity ($L_{\text{IR}}$) and intrinsic stellar UV luminosity ($L_{\text{UV}}$) shows that the $L_{\text{IR}}$ peak is typically one to 2 orders of magnitude higher than $L_{\text{UV}}$. For ultra-luminous infrared galaxies (ULIRGs), the comparison is even more extreme, with a ratio between the total luminosities exceeding $L_{\text{IR}}/L_{\text{UV}} \gg 1000$. ULIRGs exhibit a SFR comparable  to those we expect in massive ETGs \citep[e.g.,][]{Zhang+2018}. Given the similarities between these environments, their dust reprocessing efficiency is likely comparable as well.

Further research is necessary to explore the properties of the photon field generated during the formation of massive ETGs and to evaluate its consistency with observational constraints on the CMB. 
This analysis necessarily had to take assumptions and make simplifications. However, our approach builds on empirically validated models that are well supported by observations. We draw from the analysis of \citet{Eappen+2022}, from which we infer the redshift of formation of the most massive ETGs. On this foundation, we incorporate two observationally and theoretically robust frameworks, namely downsizing \citep{Thomas+2002, Kroupa+2020} and the integrated galaxy-wide IMF \citep[IGIMF, ][]{KroupaWeidner2003, Kroupa+2013, Jerabkova+2018, Yan+2021, Haslbauer+2024, Zonoozi+2025, Kroupa+2024}. The key contribution of this work is the synthesis of these established pillars with early formation redshifts. We urge the community to critically reassess the potential contribution of high-redshift foreground sources to CMB contamination.

This analysis must take into account the significant complications in data reduction associated with isolating the CMB properties (Sec.~\ref{sec:discussion}).
We note that this analytic model is a proof-of-concept for future detailed analyses that implement thorough modeling of galaxy, luminosity, and dust evolution. We stress two factors concerning this analysis: average redshift of formation, $\langle r_{\rm f} \rangle$, and average separation, $\langle d_0 \rangle$, depend on each other (Eq.~\ref{eq:n0z}). Furthermore, the epoch of formation, currently set to $15 < z < 20$, may be refined in the future due to the stringent dependence of the age-redshift relation on the chosen cosmological model.
In the view of the results documented here, it may become necessary to consider cosmological models other than the flat-$\Lambda$CDM paradigm.

\section*{Acknowledgements}
E.G. thanks Zhi-Yu Zhang (\begin{CJK}{UTF8}{gbsn}张智昱\end{CJK}) for the support and helpful discussions. We are thankful to Roland Diehl for his valuable feedback and insightful comments, and we are grateful to the anonymous referee for the constructive report and helpful suggestions. E.G. acknowledges the support of the National Natural Science Foundation of China (NSFC) under grants NOs. 12173016, 12041305.
E.G. acknowledges the science research grants from the China Manned Space Project with NOs. CMS-CSST-2021-A08, CMS-CSST-2021-A07.
E.G. acknowledges the Program for Innovative Talents, Entrepreneur in Jiangsu.
P.K. acknowledges the DAAD Eastern European Exchange program at Bonn and Charles Universities for support.

\bibliographystyle{elsarticle-harv} 
\bibliography{bibCMB-ETG}

\end{document}